\documentclass[12pt,english,draftcls, one column]{IEEEtran}
\usepackage[T1]{fontenc}
\usepackage[latin9]{inputenc}
\usepackage{amsthm}
\usepackage{amsmath}
\usepackage{amssymb}
\usepackage{graphicx}

\makeatletter
\theoremstyle{plain}
\newtheorem{thm}{\protect\theoremname}
\theoremstyle{plain}
\newtheorem{prop}[thm]{\protect\propositionname}
\theoremstyle{remark}
\newtheorem{rem}[thm]{\protect\remarkname}

\ifCLASSINFOpdf
\else
\fi
\usepackage{babel}\providecommand{\propositionname}{Proposition}
\providecommand{\remarkname}{Remark}
\providecommand{\theoremname}{Theorem}

\usepackage{babel}\providecommand{\propositionname}{Proposition}
\providecommand{\remarkname}{Remark}
\providecommand{\theoremname}{Theorem}

\makeatother

\usepackage{babel}
\providecommand{\propositionname}{Proposition}
\providecommand{\remarkname}{Remark}
\providecommand{\theoremname}{Theorem}

\begin{document}

\title{Interactive Joint Transfer of Energy and Information}

\author{P. Popovski, A. M. Fouladgar and O. Simeone 
\thanks{P. Popovski is with the Department of Electronic Systems Aalborg University,
Denmark (e-mail: petarp@es.aau.dk). A.M. Fouladgar and O. Simeone
are with the CWCSPR, New Jersey Institute of Technology, Newark, NJ
07102 USA (e-mail: \{af8,osvaldo.simeone\}@njit.edu). Part of this
paper was presented at the IEEE Information Theory Workshop (ITW 2012),
Lausanne, Switzerland, Sept. 2012.%
}
}

\maketitle





\begin{abstract}
In some communication networks, such as passive RFID systems, the
energy used to transfer information between a sender and a recipient
can be reused for successive communication tasks. In fact, from known
results in physics, any system that exchanges information via the
transfer of given physical resources, such as radio waves, particles
and qubits, can conceivably reuse, at least part, of the received
resources. 

This paper aims at illustrating some of the new challenges that arise
in the design of communication networks in which the signals exchanged
by the nodes carry both information and energy. To this end, a baseline
two-way communication system is considered in which two nodes communicate
in an interactive fashion. In the system, a node can either send an
\textquotedblleft{}on\textquotedblright{} symbol (or \textquotedblleft{}1\textquotedblright{}),
which costs one unit of energy, or an \textquotedblleft{}off\textquotedblright{}
signal (or \textquotedblleft{}0\textquotedblright{}), which does not
require any energy expenditure. Upon reception of a \textquotedblleft{}1\textquotedblright{}
signal, the recipient node \textquotedblleft{}harvests\textquotedblright{},
with some probability, the energy contained in the signal and stores
it for future communication tasks. Inner and outer bounds on the achievable
rates are derived. Numerical results demonstrate the effectiveness
of the proposed strategies and illustrate some key design insights.

\textit{Index Terms\textemdash{}} Two-way channel, interactive communication,
energy transfer, energy harvesting.
\end{abstract}



\IEEEpeerreviewmaketitle

\section{Introduction}

The conventional assumption made in the design of communication systems
is that the energy used to transfer information between a sender and
a recipient cannot be reused for future communication tasks. There
are, however, notable exceptions. An example is given by communication
based on wireless energy transfer, such as passive RFID systems \cite{RFID}
or some body area networks \cite{zhang et al}, in which a terminal
can transfer both information and energy via the transmitted radio
signal, and the delivered energy can be used for communication by
the recipients. For instance, a passive RFID tag modulates information
by backscattering the radio energy received from the reader (see,
e.g., \cite{RFID}). Another, less conventional, example is that of
a biological system in which information is communicated via the transmission
of particles (see, e.g., \cite{Eckford}), which can be later reused
for other communication tasks. A further potential instance of this
type of networks is one in which communication takes place via the
exchange of quantum systems, such as photons, which may measured and
then reused \cite{Schumacher}.

To summarize, any system that exchanges information via the transfer
of given physical resources, such as radio waves, particles or qubits,
can conceivably reuse, at least part, of the received resources for
later communication tasks. This conclusion is supported by physical
considerations \cite{Landauer} and practically demonstrated by the
existing systems based on this principle \cite{RFID,zhang et al}.
It is emphasized that the possibility to deliver jointly energy and
information promises not only to ease the energy requirements of various
communication systems, but also, more importantly, to enable novel
applications, such as the body area networks studied in \cite{RFID,zhang et al}.
Moreover, an understanding of the interplay between energy and information
flows could lead to insights on the workings of some communication
systems in nature \cite{Eckford}.

\begin{figure}[t]
\centering \includegraphics[width=9cm]{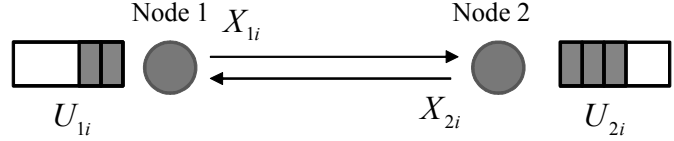} \caption{Two-way noiseless binary communication with energy exchange. The total
number of energy units is fixed (to five in the figure) and transmission
of a \textquotedbl{}1\textquotedbl{} symbol transfers energy from
the sender node to the recipient. See Fig. \ref{fig:fig1} for a generalized
model. }

\label{figsysmodel} 
\end{figure}

\subsection{State of the Art}

While the interaction between energy and information continues to
be subject of research in the physics community (see, e.g., \cite{maxwell dem}),
the topic has been tackled from a communication and information theoretic
level only in a handful of pioneering works, as reviewed in the following.
The references \cite{varshney,grover,varshney1} have focused on the
problem of maximizing the information rate of a point-to-point system
subject to minimum received energy constraints. Specifically, in \cite{varshney}
a single point-to-point channel was studied, while \cite{grover,varshney1}
investigated a set of parallel point-to-point channels. To illustrate
the trade-offs between the transfer of energy and information in a
point-to-point channel consider the noiseless transmission of a 4-PAM
signal in the alphabet $\{-2,1,1,2\}$. If one requires the received
energy to be the maximum possible, that is, to be equal to $4$, the
maximum transferable information rate is 1 bit per symbol, since one
is forced to communicate only with the larger energy symbols $\{-2,2\}.$
However, with no receive energy constraint, one can clearly convey
2 bits per symbol by choosing all available symbols with equal probability.
This example also explains the substantial difference between the
problems studied in \cite{varshney,grover,varshney1} and that with
maximum receive energy constraints studied in \cite{Gastpar}%
\footnote{With a maximal receive energy constraint of 4, one can still clearly
transmit 2 bits per symbol.%
}. The optimization of beamforming strategies under a receive energy
constraint was tackled in \cite{zhang bc,robust} for multiantenna
broadcast channels. Considerations on the design of the receiver under
the constraint that, when harvesting energy from the antenna, the
receiver is not able to use the same signal for information decoding,
can be found in \cite{switching}.

\subsection{Contributions}

In all of the previous work summarized above, the requirement on the
energy harvested from the received signal is considered to be an additional
constraint imposed to the system design. This work is instead motivated
by the observation that, in more complex network scenarios, as mentioned
above, the energy harvested from the received signal may be reused
for future communication tasks. In this case, the energy and information
content of the exchanged signals should be engineered so as to best
suit the requirements of the communication network. To study this
aspect, we consider a baseline two-way communication system, as illustrated
in Fig. \ref{figsysmodel}. This is incidentally the same topology
selected by Shannon to initiate the study of networks from an information
theoretic perspective \cite{Shannon}. In the considered model, the
two nodes interact for the exchange of information and can harvest
the received energy.

To enable analysis and insights, we assume that the two parties involved
have a common clock and that, at each time, a node can either send
an ``on\textquotedblright{} symbol (or ``1\textquotedblright{}),
which costs one unit of energy, or an ``off\textquotedblright{} signal
(or ``0\textquotedblright{}), which does not require any energy expenditure.
Upon reception of a ``1\textquotedblright{} signal, the recipient
node can harvest, possibly with some loss, the energy contained in
the signal and stores it for future communication tasks. In order
to introduce the main concepts with the minimum of the notation and
technical complications, we first consider the case in which the two
nodes start with a given number of energy units in their batteries,
which can neither be lost or replenished from outside, and the binary
channel in either direction is noiseless.

To see that even this simple scenario offers relevant research challenges,
we observe the following. If there were no limitation on the number
of energy units, the nodes could communicate 1 bit per channel use
in either direction given that the channels are ideal. However, if
there is, say, one energy unit available in the system, only the node
that currently possesses the energy unit can transmit a ``1'', whereas
the other node is forced to transmit a ``0''. Therefore, the design
of the communication strategy at the nodes should aim not only at
transferring the most information to the counterpart, but also to
facilitate energy transfer to enable communication in the reverse
direction. We study this problem, described in Sec. \ref{sec:SysModel}
by deriving inner and outer bounds on the achievable rate region as
a function of the available energy units in Sec. \ref{sec:Inner-and-Outer}.
The main results are then extended to a model that accounts for energy
replenishments and losses, along with noisy channels. The generalized
model is presented in Sec. \ref{sec:Including-Replenishments-and}
and the generalized results are presented in Sec. \ref{sec:Generalizing-the-Main}.

It is finally observed that the class of problems at hand, in which
terminals can harvest energy from the received signals is related
to the increasing body of work on energy harvesting (see, e.g., \cite{harvesting}
and references therein). However, in this line of work, the energy
is assumed to be harvested from the environment in a way that is not
affected by the communication process, unlike the scenario under study.

\emph{Notation}: $[m,n]=\{m,m+1,...,n\}$ for integers $m\leq n$.
$\mathbb{N}$ is the set of integer numbers. We use the standard notation
in \cite{El Gamal Kim} for information theoretic quantities such
as entropy and mutual information. If the distribution is $\verb"Bern"(p)$
we will also write $H(p)$ for the entropy. Capital letters denote
random variables and the corresponding lowercase quantities denote
specific values of the random variables. $X^{i}$ for an integer $i$
denotes the vector $X^{i}=(X_{1},...,X_{i})$.

\section{System Model}

\label{sec:SysModel}

We consider the binary and noiseless two-way system illustrated in
Fig. 1, in which the total number of \emph{energy units} in the system
is equal to a finite integer number $\verb"U"\geq1$ at all times
and the channels between the two nodes are noiseless. In Sec. \ref{sec:Including-Replenishments-and},
the model will be extended to include stochastic stochastic energy
losses and replenishments along with noisy channels. At any given
time instant $k$, with $k\in[1,n]$, the \emph{state} of the system
$(U_{1,k},U_{2,k})\in\mathbb{N}^{2}$ is given by the current energy
allocation between the two nodes. Specifically, a state $(U_{1,k},U_{2,k})$
indicates that at the $k$th channel use there are $U_{j,k}$ energy
units at Node $j$, with $j=1,2$. Since we assume here that $U_{1,k}+U_{2,k}=\verb"U"$
for each channel use $k\in[1,n]$ (i.e., no energy losses occur),
then, in this section and in the next, we will refer to $U_{1,k}$
as the state of the system, which always imply the equality $U_{2,k}=\verb"U"-U_{1,k}$.

At any channel use $k\in[1,n]$, each Node $j$ can transmit either
symbol $X_{j,k}=0$ or symbol $X_{j,k}=1$, and transmission of a
``1'' costs one energy unit, while symbol ``0'' does not require
any energy expenditure. Therefore, the available transmission alphabet
for Node $j$, $j=1,2$ during the $k$th channel use is\begin{subequations}\label{eq:alphabets}
\begin{eqnarray}
\mathcal{X}_{u}=\{0,1\} & \textrm{ if } & U_{j,k}=u\geq1\label{eq:}\\
\textrm{ and }\mathcal{X}_{0}=\{0\} & \textrm{ if } & U_{j,k}=0,
\end{eqnarray}
 \end{subequations}so that $X_{j,k}\in\mathcal{X}_{u}$ if $U_{j,k}=u$
energy units are available at Node $j$. The channel is noiseless
so that the received signals at channel use $k$ are given by 
\begin{equation}
Y_{1,k}=X_{2,k}\textrm{ and \ensuremath{Y_{2,k}=}\ensuremath{X_{1,k}}}
\end{equation}
 for Node 1 and Node 2, respectively.

Transmission of a ``1'' transfers one energy unit from the sender
node to the recipient node. Therefore, the state of Node 1 for $k\in[1,n]$
evolves as follows 
\begin{equation}
U_{1,k}=(U_{1,k-1}-X_{1,k-1})+X_{2,k-1},
\end{equation}
 where we set $U_{1,1}=u_{1,1}\leq\verb"U"$ as some initial state
and $U_{2,k}=\verb"U"-U_{1,k}$. We observe that the current state
$U_{1,k}$ is a deterministic function of the number $\verb"U"$ of
total energy units, of the initial state $U_{1,1}$ and of the previously
transmitted signals $X_{1}^{k-1}$ and $X_{2}^{k-1}$. We also note
that both nodes are clearly aware of the state of the system at each
time since $U_{1,k}+U_{2,k}=\verb"U"$ is satisfied for each channel
use $k$.

Node 1 has message $M_{1}$, uniformly distributed in the set $[1,2^{nR_{1}}]$,
to communicate to Node 2, and similarly for the message $M_{2}\in[1,2^{nR_{2}}]$
to be communicated between Node 2 and Node 1. Parameters $R_{1}$
and $R_{2}$ are the transmission rates in bits per channel use (c.u.)
for Node 1 and for Node 2, respectively. We use the following definitions
for an $(n,R_{1},R_{2},\verb"U")$ code. Specifically, the code is
defined by: the overall number of energy units $\verb"U"$; two sequences
of encoding functions, namely, for Node 1, we have functions $\mathrm{f}_{1,k}$
for $k\in[1,n]$, which map the message $M_{1}$ and the past received
symbols $\ensuremath{X_{2}^{k-1}}$ (along with the initial state)
into the currently transmitted signal $X_{1,k}\in\mathcal{X}_{U_{1,k}}$;
similarly, for Node 2, we have functions $\mathrm{f}_{2,k}$ for $k\in[1,n]$,
which map the message $M_{2}$ and the past received symbols $\ensuremath{X_{1}^{k-1}}$
(along with the initial state) into the currently transmitted signal
$X_{2,k}\in\mathcal{X}_{U_{2,k}}$; and two decoding functions, namely,
for Node 1, we have a function $\mathrm{g}_{1}$, which maps all received
signals $X_{2}^{n}$ and the local message $M_{1}$ into an estimate
$\hat{M}_{1}$ of message $M_{2}$; and similarly, for Node 2, we
have a function $\mathrm{g}_{2}$, which maps all received signals
$X_{1}^{n}$ and the local message $M_{2}$ into an estimate $\hat{M}_{1}$
of message $M_{1}$.

We say that rates ($R_{1},R_{2}$) are achievable with $\verb"U"$
energy units if there exists an $(n,R_{1},R_{2},\verb"U")$ code for
all sufficiently large $n$ that guarantees reliable communication.
We are interested in studying the closure of the set of all the rate
pairs $(R_{1},R_{2})$ that are achievable with $\verb"U"$ energy
units, which we refer to as capacity region $\mathcal{C}(\verb"U")$.
Given the noiseless nature of the channels, we note that the initial
state $U_{1,1}=u_{1,1}\leq\verb"U"$ does not affect the rate region
since in a finite number of steps it is always possible to redistribute
the energy according to any desired state.

\section{Inner and Outer Bounds\label{sec:Inner-and-Outer}}

In this section, we derive inner and outer bounds to the capacity
region.

\subsection{Inner Bounds}

In order to gain insights into the nature of the problem under study,
we consider here various communication strategies. We start by the
simplest, but intuitively important, case with \verb"U"$=1$, and
we then generalize to \verb"U"$>1$.

\subsubsection{\texttt{U}$=1$ Energy Unit}

We start with the special case of one energy unit ($\verb"U"=1$)
and assume the initial state $u_{1,1}=1$, so that the energy unit
is initially available at Node 1. The other case, namely $u_{1,1}=0$,
can be treated in a symmetric way. In this setting, during each channel
use, ``information'' can be transferred only from the node where
the energy unit resides towards the other node, and not vice versa,
since the other node is forced to transmits the ``0'' symbol. This
suggests that, when $\verb"U"=1$, the channel is necessarily used
in a time-sharing manner, and thus the sum-rate is at most one bit
per channel use. 
The first question is whether the sum-rate of 1 bit/c.u. is achievable,
and, if so, which strategy accomplishes this task.

\subsubsection*{A Na\"{i}ve Strategy}

We start with a rather na\"{i}ve encoding strategy that turns out
to be insufficient to achieve the upper bound of 1 bit/c.u.. The nodes
agree on a frame size $F=2^{b}>1$ channel uses for some integer $b$
and partition the $n$ channel uses in $n/F$ frames (assumed to be
an integer for simplicity). The node that has the energy unit at the
beginning of the frame communicates $b=\log_{2}F$ bits to the other
node by placing the energy unit in one specific channel use among
the $F=2^{b}$ of the frame. This process also transfers the energy
unit to the other node, and the procedure is repeated. The sum-rate
achieved by this scheme is 
\begin{equation}
R_{1}+R_{2}=\frac{\log_{2}F}{F}\textrm{ [bits/c.u.]},
\end{equation}
 which is rather inefficient: the maximum is achieved with $F=2$,
leading to a sum-rate of $R_{1}+R_{2}=1/2$ bits/ c.u..

The previous strategy can be easily improved by noting that the frame
can be interrupted after the channel use in which the energy unit
is used, since the receiving node can still decode the transmitted
$b$ bits. This strategy corresponds to using a variable-length channel
code. Specifically, we can assign, without loss of optimality within
this class of strategies, the codeword ``01'' to information bit
``0'' and the codeword ``1'' to bit ``1''. The average number
of channel uses per bit is thus $1/2+1/2$$\cdot2=3/2$ . Therefore,
the overall number of channel uses necessary for the transmission
of $m$ bits is upper bounded by $\frac{3m}{2}+m\epsilon$ with arbitrarily
small probability for large $m$ by the weak law of large numbers
(see, e.g., \cite{El Gamal Kim}). It follows that an achievable sum-rate
is given by 
\begin{equation}
R_{1}+R_{2}=\frac{1}{3/2}=\frac{2}{3},\label{eq:RateOptimizedFrame}
\end{equation}
 which is still lower than the upper bound of 1 bit/c.u..

\subsubsection*{An Optimal Strategy}

We now discuss a strategy that achieves the upper bound of $1$ bit/c.u..
The procedure is based on time-sharing, as driven by the transfer
of the energy unit from one to the other node. Specifically, each
Node $j$ has $m$ bits of information $b_{j,1},...,b_{j,m}$. If
$m$ is large enough, from the law of large numbers, approximately
half the bits will be zeros and the other half will be ones. Since
the initial state is $u_{1,1}=1$, Node 1 is the first to transmit:
it sends its information bits, starting with $b_{1,1}$ up until the
first bit that equals ``1''. Specifically, assume that we have $b_{1,1}=b_{1,2}=\cdots b_{1,i_{1}-1}=0$
and $b_{1,i_{1}}=1$. Thus, in the $i_{1}$th channel use the energy
unit is transferred to Node 2. From the $(i_{1}+1)-$th channel use,
Node 2 then starts sending its first bit $b_{2,1}$ and the following
bits until the first bit equal to ``1''. The process is then repeated.
It is easy to see that the total time required to finalize this two-way
communication is constant and equal%
\footnote{This equality is approximate if the sequences $b_{j,1},...,b_{j,m}$
do not contain exactly half zeros and half ones.%
} to $2m$ and thus the achieved sum-rate is equal to $R_{1}+R_{2}=1$
bit/c.u..

\subsubsection{$\texttt{U}>1$ Energy Units}

In the sum-capacity strategy discussed above with $\texttt{U}=1$
energy unit, both nodes transmit equiprobable symbols ``0'' and
``1''. When there are $\verb"U">1$ energy units in the system,
maximizing the sum-capacity generally requires a different approach.
Consider the scenario with $\verb"U"=2$ energy units: now it can
happen that both energy units are available at one node, say Node
1. While Node 1 would prefer to transmit equiprobable symbols ``0''
and ``1'' in order to maximize the \emph{information} flow to the
recipient, one must now also consider the \emph{energy }flow: privileging
transmission of a ``1'' over that of a ``0'' makes it possible
to transfer energy to Node 2, leading to a state in which both nodes
have energy for the next channel use. This might be beneficial in
terms of achievable sum-rate.

Based on this insight, in the following, we propose a coding strategy
that employs rate splitting and codebook multiplexing. The strategy
is a natural extension of the baseline approach discussed above for
the case $\verb"U"=1$. Each Node $j$ constructs $\verb"U"$ codebooks,
namely ${\cal C}_{j|u}$, with $u\in[1,\verb"U"]$, where codebook
${\cal C}_{j|u}$ is to be used when the Node $j$ has $u$ energy
units. Each codebook ${\cal C}_{j|u}$ is composed of codewords that
all have a specific fraction $p_{1|u}$ of ``1'' symbols. The main
idea is that, when the number $u$ of available energy units is large,
one might prefer to use a codebook with a larger fraction $p_{1|u}$
of ``1'' symbols in order to facilitate energy transfer.


\begin{prop}
\label{prop:achievable} The rate pair $(R_{1},R_{2})$ satisfying
\begin{eqnarray}
R_{1} & \leq & \sum_{u=1}^{\verb"U"}\pi_{u}H({X_{1|u})}\nonumber \\
\textrm{and }R_{2} & \leq & \sum_{u=1}^{\verb"U"}\pi_{u}H({X_{2|u})},\label{eq:Rate}
\end{eqnarray}
where $X_{j|u}\sim\verb"Bern"(p_{j|u})$, $j=1,2$, for some probabilities
$0<p_{1|u},p_{2|u}<1$, $u=1\ldots\verb"U"$, with $p_{1|0}=p_{2|\verb"U"}=0$,
is included in the capacity region $\mathcal{C}(\verb"U")$. The probabilities
$\pi_{u}\geq0,\textrm{ }u=0\ldots\verb"U"$, in (\ref{eq:Rate}) satisfy
the fixed-point equations 
\begin{align}
\pi_{u}=\pi_{u}(\phi_{0,0|u}+\phi_{1,1|u})+\pi_{u-1}\phi_{0,1|u}+\pi_{u+1}\phi_{1,0|u}\label{eq:MarkovChainTransitions}
\end{align}
 with $\pi_{-1}=\pi_{\verb"U"+1}=0$, $\sum_{u=1}^{\verb"U"}\pi_{u}=1$,
and we have defined 
\begin{eqnarray}
\phi_{0,0|u} & = & (1-p_{1|u})(1-p_{2|\verb"U"-u})\nonumber \\
\phi_{0,1|u} & = & (1-p_{1|u})p_{2|\verb"U"-u}\nonumber \\
\phi_{1,0|u} & = & p_{1|u}(1-p_{2|\verb"U"-u})\nonumber \\
\textrm{and }\phi_{1,1|u} & = & p_{1|u}p_{2|\verb"U"-u}.\label{eq:trans_prob}
\end{eqnarray}

\end{prop}
\noindent This proposition is proved by resorting to random coding
arguments, whereby codebook ${\cal C}_{j|u}$ is generated with independent
and identically distributed (i.i.d.) entries $X_{j|u}$ distributed
as $\verb"Bern"(p_{j|u})$, $j=1,2$. As introduced above, the idea
is that, when the state is $U_{1,i}=u$, Node $j$ transmits a symbol
from the codebook associated with that state, namely codebook ${\cal C}_{1|u}$
for Node $1$ and codebook $\mathcal{C}_{2|\verb"U"-u}$ for Node
2 (which has $\verb"U"-u$ energy units). Both nodes know the current
state $U_{1,i}$ and thus can demultiplex the codebooks at the receiver
side. According to the random coding argument, the state $U_{1,i}$
evolves according to a Markov chain: the system stays in the same
state $u$ with probability $\phi_{0,0|u}+\phi_{1,1|u}$ (both nodes
transmit ``0'' or ``1''), changes to the state $u+1$ with probability
$\phi_{1,0|u}$ (Node 1 transmits a ``1'' and Node 2 a ``0'')
or changes to the state $u-1$ with probability $\phi_{0,1|u}$ (Node
1 transmits a ``0'' and Node 2 a ``1''). The definition of the
conditional probabilities (\ref{eq:trans_prob}) reflects the fact
that the codebooks are generated independently by the two nodes. A
full proof is given in Appendix~\ref{sec:ProofProp1}.

\subsection{Outer Bounds}

\label{sec:Converse}

In this section, we derive an outer bound to the capacity region $\mathcal{C}(\verb"U")$.
Similar to the standard cut-set bound \cite[Ch. 17]{El Gamal Kim},
the outer bound differs from the inner bound of Proposition \ref{prop:achievable}
in that it allows for a joint distribution $\phi_{x_{1},x_{2}|u}$
of the variables $X_{1|u}$ and $X_{2|u}$. 
\begin{prop}
If the rate pair ($R_{1},R_{2}$) is included in the capacity region
$\mathcal{C}(\verb"U")$, then there exist probabilities $\pi_{u}\geq0$
with $\sum_{u=1}^{\verb"U"}\pi_{u}=1$, and \textup{$\phi_{x_{1},x_{2}|u}\geq0$
with }$\sum_{x_{1},x_{2}\in\{0,1\}}\phi_{x_{1},x_{2}|u}=1$ for all
$u\in\{0,1,...,\verb"U"\}$, such that $\phi_{1,x_{2}|0}=0$ for $x_{2}\in\{0,1\}$,
$\phi_{x_{1},1|\verb"U"}=0$ for $x_{1}\in\{0,1\}$, condition (\ref{eq:MarkovChainTransitions})
is satisfied, and the following inequalities hold 
\begin{align}
R_{1} & \leq\sum_{u=0}^{U}\pi_{u}H\left(X_{1|u}\mid X_{2|u}\right)\\
R_{2} & \leq\sum_{u=0}^{U}\pi_{u}H\left(X_{2|u}\mid X_{1|u}\right)\\
\textrm{and }R_{1}+R_{2} & \leq\sum_{u=0}^{U}\pi_{u}H\left(X_{1|u},X_{2|u}\right),\label{eq:sum_rate_upper}
\end{align}
where variables $X_{1|u}$ and $X_{2|u}$ are jointly distributed
with distribution $\phi_{x_{1},x_{2}|u}$. 
\end{prop}
\noindent The outer bound above can be interpreted as follows. Suppose
that, when the state is $U_{1,k}=u$, the nodes were allowed to choose
their transmitted symbols according to a \emph{joint} distribution
$\phi_{x_{1},x_{2}|u}=\Pr[X_{1,k}=x_{1},\textrm{ }X_{2,k}=x_{2}]$.
Note that this is unlike the achievable strategy described in the
previous section in which the codebook were generated independently.
Intuitively, allowing for correlated codebooks, leads to a larger
achievable rate region, as formalized by Proposition 2, whose proof
can be found in Appendix B.

\subsection{Numerical Results}

Fig.~\ref{fig:num_results} compares the achievable sum-rate obtained
from Proposition 1 and the upper bound (\ref{eq:sum_rate_upper})
on the sum-rate obtained from Proposition 2 versus the total number
of energy units $\verb"U"$. As for the achievable sum-rate, we consider
both a conventional codebook design in which the same probability
$p_{j|u}=0.5$ is used irrespective of the state $U_{1.i}=u$, and
one in which the probabilities $p_{j|u}$ are optimized. It can be
seen that using conventional codebooks, which only aim at maximizing
information flow on a single link, leads to substantial performance
loss. Instead, the proposed strategy with optimized probabilities
$p_{j|u}$, which account also for the need to manage the energy flow
in the two-way communication system, performs close to the upper bound.
The latter is indeed achieved when $\verb"U"$ is large enough.

A remark on the optimal probabilities $p_{j|u}$ is in order. Due
to symmetry, it can be seen that we have $p_{1|u}=p_{2|\verb"U"-u}$.
Moreover, numerical results show that $p_{1|u}$ increases monotonically
as $u$ goes from $0$ to \verb"U", such that $p_{1,\verb"U"}>0.5$.
In particular, when the number of states $\verb"U"+1$ is odd, it
holds that $p_{1,\verb"U"/2}=p_{2,\verb"U"/2}=0.5$. It is finally
noted that the energy neutral transitions (both nodes emitting ``0''
or both emitting ``1'') occur with equal probability (i.e., $(1-p_{1,u})(1-p_{2,u})=p_{1,u}p_{2,u}$).

\begin{figure}[t]
\centering \includegraphics[width=9cm]{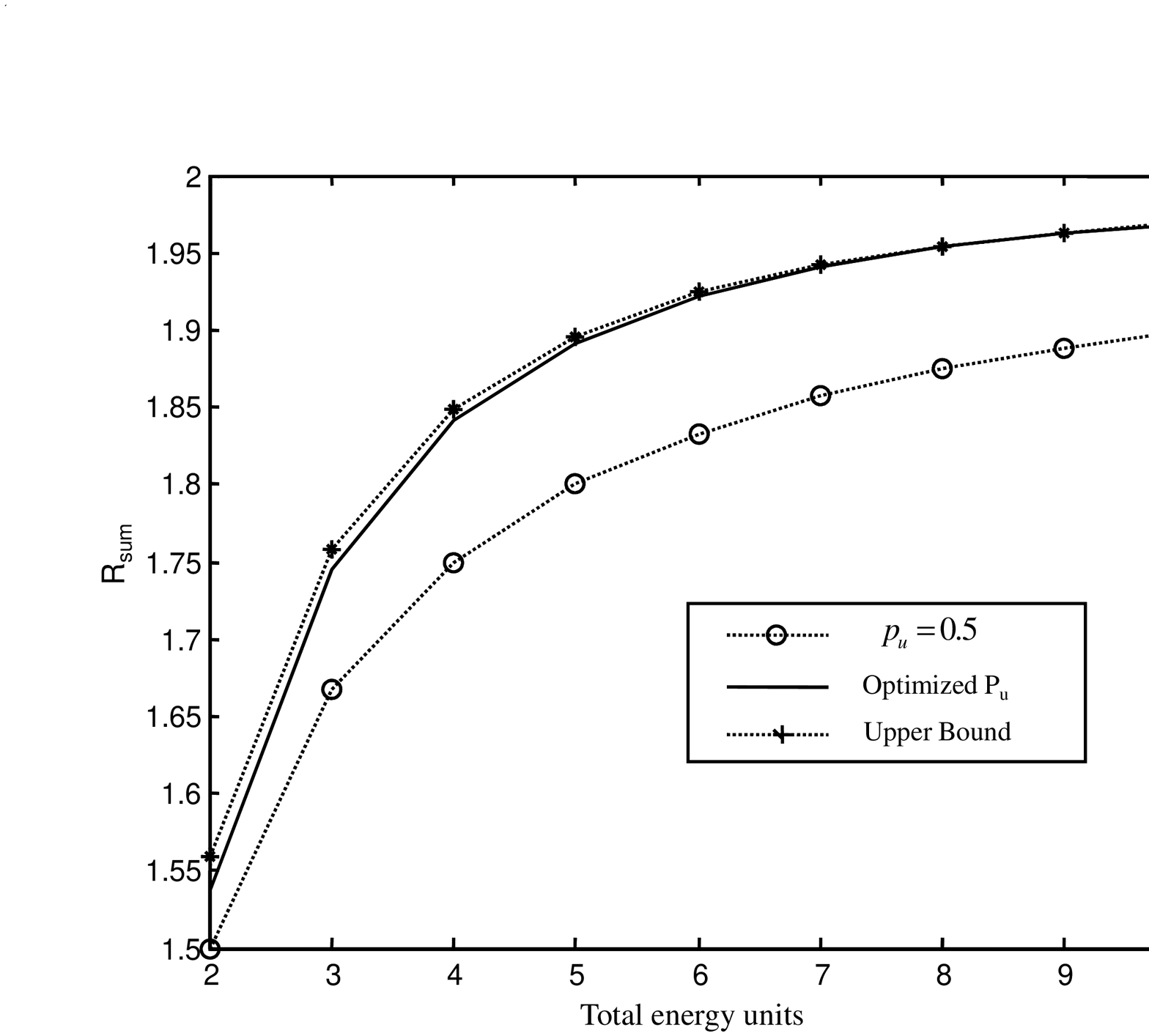} \caption{Achievable sum-rate obtained from Proposition 1 and upper bound (\ref{eq:sum_rate_upper})
versus the total number of energy units \texttt{U}. }

\label{fig:num_results} 
\end{figure}

\section{System Model with Stochastic Replenishments and Losses\label{sec:Including-Replenishments-and}}

\begin{figure}[h!]
\centering \includegraphics[bb=3bp 551bp 768bp 733bp,clip,width=9cm]{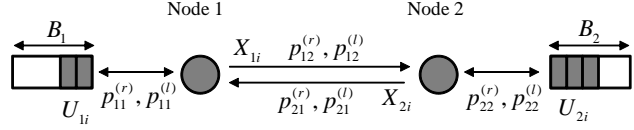}
\caption{Two-way noisy binary communication with energy exchange. The probabilities
of replenishments through the channel or locally at the nodes are
referred to as $p_{ij}^{(r)},\textrm{ with }i\neq j$ or $p_{ij}^{(r)},\textrm{ with }i=j$,
respectively, and similarly for the probabilities of losses $p_{ij}^{(l)}$.
See Fig. \ref{fig:fig2} for an illustration of the channel and Fig.
\ref{fig:fig3} for an illustration of the harvesting process.}

\label{fig:fig1} 
\end{figure}

In this section we extend the two-way communication system with energy
exchange studied above to include energy losses and replenishments,
which may occur in different parts of the system, as illustrated in
Fig. \ref{fig:fig1}. Specifically, the energy units can be lost either
while in transit through a lossy channel or locally at either node
during processing. Similarly, energy units can be replenished either
by harvesting energy from the channel, e.g., from an interfering signal
or a source of RF energy, or through a source of power locally connected
to the node, e.g., a solar panel. All loss and replenishment events
are assumed to be independent. As above, we assume that the two parties
involved have a common clock, and that, at each time, a node can either
send a ``1'', which requires one unit of energy cost, or a ``0'',
which does not require any energy expenditure. We also assume that
Node 1 and Node 2 have energy buffers of capacities $B_{1}$ and $B_{2}$
energy units, respectively, to store the available energy.

\begin{figure}[h!]
\centering \includegraphics[bb=152bp 356bp 589bp 618bp,clip,scale=0.35]{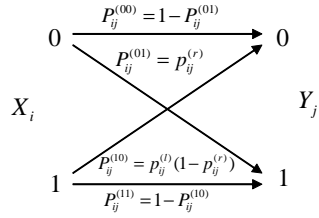}
\caption{Channel from Node $i$ to Node $j$.}

\label{fig:fig2} 
\end{figure}

Unlike in the previous sections, we assume that the binary channel
from Node $i$ to Node $j$ with $i\neq j$, is noisy as shown in
Fig. \ref{fig:fig2}, with the probability of $P_{ij}^{(01)}$ of
flipping a ``0'' symbol to a ``1'' symbol and the probability
$P_{ij}^{(10)}$ of flipping symbol ``1'' to symbol ``0''. These
probabilities can be interpreted in terms of replenishments and losses
across the channel. To elaborate, let us define as $p_{ij}^{(r)}$
the probability of replenishment via harvesting from the channel (for
$i\neq j$), e.g., thanks to an RF source that operates on the same
bandwidth as the $ij$ link. Moreover, define as $p_{ij}^{(l)}$ the
probability that an energy unit is lost while in transmit through
the channel for the $ij$ link. With these definitions, assuming that
losses and replenishments are independent, we can write the transition
probabilities as in Fig. \ref{fig:fig2}.

\begin{figure}[h!]
\centering \includegraphics[bb=152bp 356bp 589bp 618bp,clip,scale=0.35]{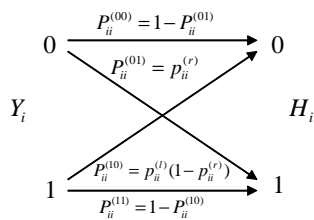}
\caption{Statistical relationship between the received signal $Y_{i}$ and
the energy $H_{i}$ harvested by Node $i$. }

\label{fig:fig3} 
\end{figure}

Losses and replenishments can also take place locally at the nodes
with the probability $P_{ii}^{(01)}$ of flipping a ``0'' symbol
to a ``1'' symbol and the probability $P_{ii}^{(10)}$ of flipping
symbol ``1'' to symbol ``0'' at Node $i$ upon reception. Specifically,
let us define as $p_{ii}^{(r)}$ the probability of replenishment
at Node $i$, whereby an energy unit is received by Node $i$ from
an external source of energy directly connected to the node, such
as a solar panel. Note that this energy unit is not received through
the channel but is directly stored in the buffer and therefore does
not affect the decoder, unlike replenishment events over the channel.
Moreover, define as $p_{ii}^{(l)}$ the probability that an energy
unit, while correctly received by the decoder at Node $i$, is lost
during processing before reaching the energy buffer. Note that in
this case the decoder at Node $i$ correctly records a \textquotedbl{}1\textquotedbl{},
but this energy unit cannot be reused for future channel uses. This
event is thus different from a loss over the channel in which the
decoder at Node $i$ observes a \textquotedbl{}0\textquotedbl{} symbol.
With these definitions, assuming that losses and replenishments are
independent and that no more than one energy unit can be harvested
in each time instant, we can write the transition probabilities between
the received signal $Y_{i}$ and the harvested energy $H_{i}$ at
Node $i$ as in Fig. \ref{fig:fig3}. 

Based on the discussion above, at any given time instant $k$, with
$k\in\left[1,n\right]$, the \textit{state }of the system $(U_{1,k},U_{2,k})\in\mathbb{N}^{2}$
is given by the current energy levels $U_{1,k}$ and $U_{2,k}$ in
the buffers of Node 1 and Node 2, respectively. By the capacity limitations
of the buffers, we have the inequalities $u_{1}\in[0,B_{1}]$ and
$u_{2}\in[0,B_{2}]$ for each channel use $k\in\left[1,n\right]$.
The transmitted symbols are limited as per (\ref{eq:alphabets}).

The channel is noisy with transition probabilities as in Fig. \ref{fig:fig1}.
Moreover, the relationship between received signal and harvested energy
is as in Fig. \ref{fig:fig3}. Therefore, the state of battery at
Node 1 for $k\in\left[1,n\right]$ evolves as follows 
\begin{equation}
U_{i,k}=(U_{i,k-1}-X_{i,k-1})+H_{i,k-1}.
\end{equation}

Similar to Sec. II, we use the following definitions for an $(n,R_{1},R_{2},B_{1},B_{2})$.
Specifically, the code is defined by: the buffer capacities $B_{1}$
and $B_{2}$; two sequences of encoding functions, $f_{i,k}$ for
$k\in\left[1,n\right]$ and $i=1,2$, which map the message $M_{i}$,
the past received symbols $Y_{i}^{k-1}$ along with the past and current
states ($U_{1}^{k},U{}_{2}^{k}$) into the currently transmitted signal
$X_{i,k}\in\mathcal{X}_{U_{i,k}}$; two decoding functions $g_{i}$,
for $i=1,2$, which map the received signal $Y_{i}^{n}$, the local
message $M_{i}$ and the sequence of states $U{}_{1}^{n},U{}_{2}^{n}$
into an estimate $\hat{M}_{j}$ of message $M_{j}$ of the other node
$j\neq i$. Achievability is defined as in Sec. II. Finally, the closure
of the set of all the rate pairs $(R_{1},R_{2})$ is defined as the
capacity region $\mathcal{C}_{0}(B_{1},B_{1})$, where subscript ``0''
denotes the fact that the capacity region generally depends on the
initial state. 
\begin{rem}
\label{rem:In-the-definition}In the definition of code given above,
we have assumed that the nodes can track the state of the batteries
$(U_{1,k},U_{2,k})$ at both nodes. We refer to this scenario as having
Global Energy Information (GEI). We remark that in the presence of
losses and replenishment, the nodes generally cannot track the amount
of energy available at the other node based only on the knowledge
of the received signal. Instead, information about the state of the
other node needs to be acquired through additional resources such
as control channels or appropriate sensors. In general, the assumed
model with GEI can thus be thought of providing a best-case bound
on system performance. In Sec. \ref{sub:Local-Energy-Information},
we will study the scenario, referred to as having Local Energy Information
(LEI), in which each node is only aware of the energy available in
its own local battery. 
\end{rem}

\section{Generalizing the Inner and Outer Bounds\label{sec:Generalizing-the-Main}}

In this section, we first propose a communication strategy that leads
to an achievable rate region by generalizing the approach discussed
in Sec. \ref{sec:Inner-and-Outer}. The outer bound of Sec. \ref{sec:Inner-and-Outer}
is similarly extended. While the strategy at hand is based on GEI
(see Remark \ref{rem:In-the-definition}), we then discuss an achievable
strategy with LEI in Sec. \ref{sub:Local-Energy-Information}, and
present some numerical results in Sec. \ref{sub:Numerical-Results}.

\subsection{Transmission Strategy}

The proposed strategy is an extension of the approach put forth in
Sec. \ref{sec:Inner-and-Outer}, and operates as follows. Node $i$,
with $i=1,2$, constructs one independent codebook $\mathcal{C}_{i|(u_{1},u_{2})}$
for each possible sate $(u_{1},u_{2})\in[0,B_{1}]\times[0,B_{2}]$.
As in Sec. \ref{sec:Inner-and-Outer}, at each time $k$, if the state
is $(U_{1,k},U_{2,k})=(u_{1},u_{2})$, then Node $i$ transmits the
next symbol from the codebook $\mathcal{C}_{i|(u_{1},u_{2})}$. At
the end of the last channel use, each node, being aware of the sequences
of states, can demultiplex the transmission of the other node and
decode the messages encoded in all the $(B_{1}+1)(B_{2}+1)$ codebooks.

The codebook of Node $i$ corresponding to state $(u_{1},u_{2})$
is generated by drawing each bit independently with a given probability
$p_{i|(u_{1},u_{2})}$ for $i=1,2$ and all states $(u_{1},u_{2})\in[0,B_{1}]\times[0,B_{2}]$.
Note that, due to (\ref{eq:alphabets}), we have $p_{1|(0,u_{2})}=0$
for all $u_{2}\in[0,B_{2}]$ since, when $U_{1,k}=0$, Node 1 has
no energy available and thus must transmit a ``0'' symbol; and similarly
we have $p_{2|(u_{1},0)}=0$ for all $u_{1}\in[0,B_{1}]$. Given the
probabilities $p_{i|(u_{1},u_{2})}$ for $i=1,2$ and all states $(u_{1},u_{2})\in[0,B_{1}]\times[0,B_{2}]$,
the $(B_{1}+1)(B_{2}+1)\times(B_{1}+1)(B_{2}+1)$ transition probability
matrix $\boldsymbol{P}$ can be obtained that contains the transition
probabilities from any state $(u_{1},u_{2})\in[0,B_{1}]\times[0,B_{2}]$
to any state $(u_{1}^{\prime},u_{2}^{\prime})\in[0,B_{1}]\times[0,B_{2}]$.
These transition probabilities depend on the parameters $(p_{ij}^{(r)},p_{ij}^{(l)})$,
$(p_{ii}^{(r)},p_{ii}^{(l)})$, and $p_{i|(u_{1},u_{2})}$ for $i,j=1,2$
and $(u_{1},u_{2})\in[0,B_{1}]\times[0,B_{2}]$, as detailed in Appendix
\ref{app:C}.

\subsection{Inner and Outer Bounds}

In order to derive the rates achievable with this strategy, denote
as $\pi_{(u_{1},u_{2})}$ the average fraction of channel uses $k$
such that we have $(U_{1,k},U_{2,k})=(u_{1},u_{2})$ for all states
$(u_{1},u_{2})\in[0,B_{1}]\times[0,B_{2}]$, as done in Sec. \ref{sec:Inner-and-Outer}.
Note that $\sum_{(u_{1},u_{2})}\pi_{(u_{1},u_{2})}=1$. This function
is also referred to as the steady-state probability and can be calculated
as the limit 
\begin{eqnarray}
\boldsymbol{\pi}_{(u_{1},u_{2})} & = & \underset{k\rightarrow\infty}{\lim}\boldsymbol{P}^{k}\boldsymbol{\pi}(1),\label{eq:eq4-1}
\end{eqnarray}
 where $\boldsymbol{\pi}_{(u_{1},u_{2})}$ is the $(B_{1}+1)(B_{2}+1)\times1$
vector containing the steady-state probabilities $\pi_{(u_{1},u_{2})}$
for all states $(u_{1},u_{2})\in[0,B_{1}]\times[0,B_{2}]$ and we
recall that $\boldsymbol{P}$ is the transition probability matrix.
Vector $\boldsymbol{\pi}(1)$ accounts for the initial state and is
thus a vector of all zeros except for a one in the entry corresponding
to the initial state. We note that the limit in (\ref{eq:eq4-1})
always exists for the model studied in Sec. \ref{sec:Inner-and-Outer}
(for all non-trivial transmission probabilities), and is given by
(\ref{eq:MarkovChainTransitions})-(\ref{eq:trans_prob}). The same
is generally true here apart from degenerate cases. However, the transition
matrix (\ref{eq:eq4-1}) is possibly reducible, and thus the calculation
of the limit generally requires the factorization of the matrix according
to the canonical form for reducible matrices. We refer to \cite[ch. 8]{Meyer}
for a detailed discussion on the existence and calculation of the
limit (\ref{eq:eq4-1}). 
\begin{prop}
Assuming that the limit (\ref{eq:eq4-1}) exists, the rate pair $(R_{1},R_{2})$
satisfying the inequalities 
\begin{eqnarray}
R_{1} & \leq & \underset{\underset{[0,B_{1}]\times[0,B_{2}]}{(u_{1},u_{2})\in}}{\sum}\pi_{(u_{1},u_{2})}I(X_{1|(u_{1},u_{2})};Y_{2})\nonumber \\
\textrm{and }R_{2} & \leq & \underset{\underset{[0,B_{1}]\times[0,B_{2}]}{(u_{1},u_{2})\in}}{\sum}\pi_{(u_{1},u_{2})}I(X_{2|(u_{1},u_{2})};Y_{1})\label{eq:eq4}
\end{eqnarray}
 for some transmission probabilities $p_{i|(u_{1},u_{2})}$, for $i=1,2$
and $(u_{1},u_{2})\in[0,B_{1}]\times[0,B_{2}]$ is achievable, where
we have denoted as $X_{i|(u_{1},u_{2})}$ as the Bernoulli variable
$\textrm{Bern}(p_{i|(u_{1},u_{2})})$. We also have 
\begin{eqnarray}
I(X_{1|(u_{1},u_{2})};Y_{2}) & = & H\left((1-p_{1|(u_{1,}u_{2})})P_{01}+p_{1|(u_{1,}u_{2})}P_{11}\right)\nonumber \\
 &  & -\left[p_{1|(u_{1,}u_{2})}H(P_{11})+(1-p_{1|(u_{1,}u_{2})})H(P_{01})\right]
\end{eqnarray}
 and similarly for $I(X_{2|(u_{1},u_{2})};Y_{1})$. \end{prop}
\begin{rem}
The achievability of the rates in (\ref{eq:eq4}) can be proved by
adopting the multiplexing strategy described above and following the
same main steps as in Appendix A. Here, we also point out that the
achievability of (\ref{eq:eq4}) under the assumption that the limit
(\ref{eq:eq4-1}) exists is a direct consequence of \cite[Lemma 12.3.1]{Gray}. 
\end{rem}
An outer bound can be also derived by generalizing Proposition 2.
In particular, following similar steps as in Appendix B, one can prove
that an outer bound is obtained by allowing for joint probabilities,
rather than product distributions as in Proposition 1. Moreover, one
can add the sum-rate constraint that generalizes (\ref{eq:sum_rate_upper})
as 
\begin{equation}
R_{1}+R_{2}\leq\underset{\underset{[0,B_{1}]\times[0,B_{2}]}{(u_{1},u_{2})\in}}{\sum}\pi_{(u_{1},u_{2})}I(X_{1|(u_{1},u_{2})},X_{2|(u_{1},u_{2})};Y_{1},Y_{2}),\label{eq:-1}
\end{equation}
 where $X_{1|(u_{1},u_{2})},X_{2|(u_{1},u_{2})}$ are jointly distributed.

\subsection{Local Energy Information\label{sub:Local-Energy-Information} }

In the discussion above, we have assumed GEI, that is, each node knows
the full current energy state $(U_{1,k},U_{2,k})$ (see Remark \ref{rem:In-the-definition}).
In this section, we consider instead the scenario with LEI, in which
Node 1 only knows its local energy level $U_{1}$ and Node 2 only
knows $U_{2}$.

We first observe that the energy $U_{1,k}$ can be considered to be
the state of the link 12 at channel use $k$, since it affects the
available input symbols via (\ref{eq:alphabets}) (and similarly for
$U_{2,k}$ and link 21). Therefore, the model at hand falls in the
category of channels with states in which the state is known only
at the transmitter. For these channels, under the assumption that
the state sequence is i.i.d. and independent of the transmitted signal,
it is known that so called Shannon strategies are optimal \cite[Ch. 7]{El Gamal Kim}.
In the model under study, unlike the conventional setting, the state
sequence $U_{1}^{n}$ (and $U_{2}^{n}$) is neither i.i.d. nor independent
of the transmitted signal $X_{1}^{n}$ (and $X_{2}^{n}$). Therefore,
Shannon strategies are generally not optimal. We will see below that
they can be nevertheless used to lead to non-trivial achievable rates.

Following Shannon strategies, we draw auxiliary codebooks made of
independent and i.i.d. codewords $V_{1}^{n}$ and $V_{2}^{n}$ using
pmfs $p(v_{1})$ and $p(v_{2})$, respectively. Each symbol $V_{j,k}$
for Node $j$ and time instant $k$ is a vector consisting of $B_{j}$
bits. The main idea is that, at each time $k$, Node $j$ transmits
the bit in $V_{j,k}$ corresponding to the current state $U_{j,k}$.
Note that the latter can take $B_{j}$ possible values at which the
transmitted signal is non-trivial (for $U_{j,k}=0,$ we necessarily
have $X_{j,k}=0$).

At the receiver side, the decoder at Node 2 uses joint typicality
decoding with respect to the distribution $p(v_{1},y_{2})$, which
is given as 
\begin{eqnarray}
p(v_{1},y_{2}) & = & p(v_{1})\underset{u_{1}}{\sum}\pi(u_{1})p(y_{2}|f_{1}(v_{1},u_{1}))\label{eq:7}
\end{eqnarray}
 where $\pi(u_{1})$ is the marginal distribution of the steady-state
probability of the Markov chain induced by the random coding strategy
and the evolution of the system, as discussed above (see also Appendix
C). Following standard information theoretic considerations, we obtain
that the rate pair $(R_{1},R_{2})$ satisfying\begin{subequations}
\begin{eqnarray}
R_{1} & \leq & I(V_{1};Y_{2}),\\
\textrm{and }R_{2} & \leq & I(V_{2};Y_{1}),\label{eq:8}
\end{eqnarray}
 \end{subequations}for some pmfs $p(v_{1})$, $p(v_{2})$ is achievable,
where $p(v_{1},y_{2})$ is as in (\ref{eq:7}) and similarly for $p(v_{2},y_{1})$.
Regarding the details of the proof, being based on conventional tools
(see \cite[Ch. 3]{El Gamal Kim}), here we simply point out that it
is based on the ergodicity of the Markov chain, which allows to conclude
that the error event in which the correct codeword is not jointly
typical takes place with negligible probability; and the packing lemma
in \cite[Lemma 3.1]{El Gamal Kim}, which entails that the error events
due to mistaking other codewords for the correct one have also negligible
probability%
\footnote{The packing lemma does not assume that the received signal be i.i.d.
and thus applies to our scenario (see \cite[Lemma 3.1]{El Gamal Kim}).%
}.
\begin{rem}
In the strategy proposed above, each node adapts the choice of the
current transmitted symbol only to the current local energy state.
A potentially better approach would be to perform adaptation based
on a local state that includes also a number of past energy states
of the node, along with the current one, and/or current and past received
signals. This aspect is not further explored in this paper. 
\end{rem}

\subsection{Numerical Results\label{sub:Numerical-Results}}

In this section, we present some numerical examples in order to assess
the impact of replenishment and loss processes. We assume that each
node has the ability to store only one unit of energy i.e., $B_{1}=B_{2}=1$,
and we consider a symmetric system with $p_{12}^{(r)}=p_{21}^{(r)}=p_{r,c}$,
$p_{11}^{(r)}=p_{22}^{(r)}=p_{r,n}$, $p_{12}^{(l)}=p_{21}^{(l)}=p_{l,c}$
and $p_{11}^{(l)}=p_{21}^{(l)}=p_{l,n}$, where the subscripts ``c''
and ``n'' stand for ``channel'' and ``node'', so that, e.g.,
$p_{r,n}$ is the probability of replenishment locally at a node.
We first assume GEI.

Fig. 5 shows the sum-rate obtained by summing the right-hand sides
of (\ref{eq:eq4}), optimized over the probabilities $p_{1|(u_{1,}u_{2})}$
and $p_{2|(u_{1,}u_{2})}$ for all states $(u_{1},u_{2})\in[0,B_{1}]\times[0,B_{2}]$
versus the replenishment probability on the channel $p_{r,c}$ (see
Fig. \ref{fig:fig2}) for two cases, namely $p_{r,n}=0,p_{l,n},p_{l,c}=0.1$
and $p_{r,n}=0,p_{l,n},p_{l,c}=0.3$. We also show in the same figure
the steady-state probability $\pi_{(1,1)}$ of state $(u_{1},u_{2})=(1,1)$
corresponding to the optimal values of $p_{1|(u_{1,}u_{2})}$ and
$p_{2|(u_{1,}u_{2})}$. It is seen that increasing the probability
$p_{r,c}$ increases the chance of being in state $(u_{1},u_{2})=(1,1)$,
due to the increased availability of energy. However, increasing $p_{r,c}$
has also the deleterious effect of flipping bits on the channel from
``0''s to ``1''s with larger probability. It is seen that, in
the regime in which $p_{r,c}$ is sufficiently small, and the system
is energy-limited, increasing $p_{r,c}$ is beneficial, while for
$p_{r,c}$ large enough the second effect dominates and the achievable
sum-rate decreases.

We now turn to assessing the effect of local replenishment at the
node. Specifically, Fig. 6 shows the optimized sum-rate versus $p_{r,n}$.
As we can see, increasing $p_{r,n}$ improves the sum-rate, since
it enhances the probability of being in state $(u_{1},u_{2})=(1,1)$,
without any side effect since it does not impair the channels.

\begin{figure}[h!]
\centering \includegraphics[clip,scale=0.35]{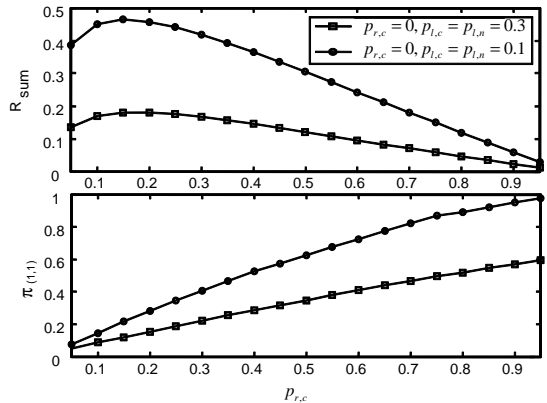} \caption{Sum-rate $R_{sum}$ and steady-state probability $\pi_{(1,1)}$ versus
the probability $p_{r,c}$ of replenishment on the channel (see Fig.
\ref{fig:fig2}).}

\label{fig:fig5} 
\end{figure}

\begin{figure}[h!]
\centering \includegraphics[clip,scale=0.35]{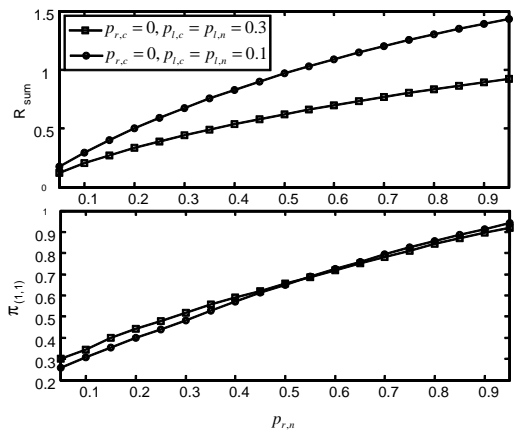} \caption{Sum-rate $R_{sum}$ and steady-state probability $\pi_{(1,1)}$ versus
the probability $p_{r,n}$ of replenishment at the node (see Fig.
\ref{fig:fig3}).}

\label{fig:fig6} 
\end{figure}

\begin{figure}[h!]
\centering \includegraphics[clip,scale=0.35]{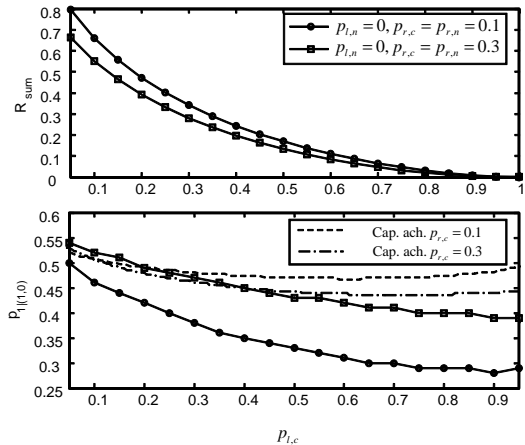} \caption{Sum-rate $R_{sum}$ and optimum transmission probability $p_{1|(1,0)}$
versus the probability $p_{l,c}$ of loss on the channel. Dotted lines
show the capacity achieving probability. }

\label{fig:fig7} 
\end{figure}

\begin{figure}[h!]
\centering \includegraphics[clip,scale=0.35]{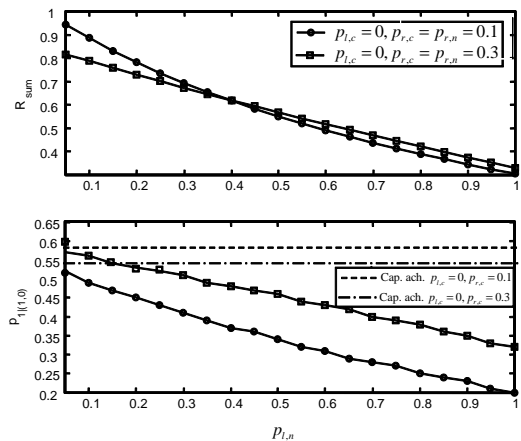} \caption{Sum-rate $R_{sum}$ and optimum transmission probability $p_{1|(1,0)}$
versus the probability $p_{l,n}$ of loss at the node. Dotted lines
show the capacity achieving probability.}

\label{fig:fig8} 
\end{figure}

Fig. \ref{fig:fig7} and Fig. \ref{fig:fig8} show the effect of loss
events on the channel and at the nodes, respectively. We show both
the sum-rate and the optimal transmission probability $p_{1|(1,0)}$,
which equals the optimal probability $p_{2|(0,1)}$ by symmetry. The
latter is also compared with the transmission probability that maximizes
the mutual information $I(X_{1|(0,1)};Y_{2})$ in (\ref{eq:eq4})
and that is thus capacity achieving. It is noted that this is the
probability that maximizes the information rate when there are no
energy limitations. As it can be seen, by comparing Fig. \ref{fig:fig7}
and Fig. \ref{fig:fig8}, increasing the loss probability both on
the channel and at the node decreases the sum-rate, although the rate
of this decrease is larger for the latter, since, similar to the discussion
above, a loss at the node does not affect the channel. Moreover, for
small $p_{l,c}$ and $p_{l,n}$, the transmission probability $p_{1|(1,0)}$
is close to the capacity-achieving probability, while for larger loss
probabilities $p_{l,c}$ and $p_{l,n}$, it becomes smaller than the
capacity-achieving probability.

\begin{figure}[h!]
\centering \includegraphics[clip,scale=0.35]{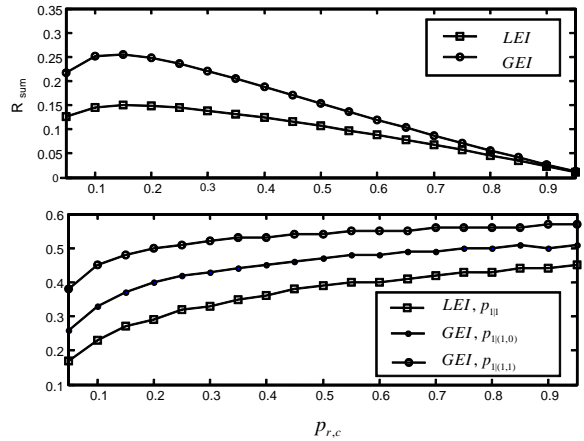} \caption{Sum-rate $R_{sum}$ and optimum transmission probabilities for Global
Energy Information (GEI) and Local Energy Information (LEI) for $p_{l,c}=0.3$
and $p_{r,n}=p_{l,n}=0$.}

\label{fig:fig9} 
\end{figure}

We now consider the effect of LEI. Fig. \ref{fig:fig9} compares the
sum-rate achieved with GEI and LEI versus the replenishment probability
$p_{r,c}$ on the channel. As it can be seen, LEI entails a significant
performance loss with respect to GEI. To gain some insight as to the
reasons of this loss, the figure also shows the optimal transmission
probabilities $p_{1|(1,0)}$, $p_{1|(1,1)}$ with GEI and the probability
$p(v_{1})=p_{1|1}$, that is the probability of transmitting \textquotedbl{}1\textquotedbl{}
if the local battery contains energy, for LEI ($V_{1}$ is a Bernoulli
variable since $B_{1}=B_{2}=1$). With GEI, the nodes can adapt the
transmission strategy to the energy state of both nodes and thus choose
different probabilities $p_{1|(1,0)}$ and $p_{1|(1,1)}$, while with
LEI the nodes are forced to choose a single probability $p_{1|1}$
irrespective of the state of the battery at the other node.

\section{Conclusions}

Energy and information content are two contrasting criteria in the
design of a communication signal. In a number of emerging and envisaged
communication networks, the participating nodes are able to reuse
part of the energy in the received signal for future communication
tasks. Therefore, it becomes critical to develop models and theoretical
insights into the involved trade-offs between energy and information
exchange at a system level. In this work, we have taken a first step
in this direction, by considering a two-way channel under a simple
binary ``on-off'' signaling model. The derived inner and outer bounds
shed light into promising transmission strategies that adapt to the
current energy state. It is emphasized that conventional strategies
based solely on the maximization of the information flow entail substantial
losses.

The results presented in this paper call for further studies on different
fronts. One is the development of better models which strike a good
balance between adherence to reality and analytical tractability.
A second is the development of better communication strategies for
the practical scenario in which the energy state of the network is
not fully known at the nodes.

\appendices{}

\section{Proof of Proposition 1\label{sec:ProofProp1}}

\subsubsection{Code construction}

We generate $\verb"U"$ codebooks for each Node $j=1,2$, namely ${\cal C}_{j|u}$,
with $u\in[1,\verb"U"]$. The codebook ${\cal C}_{j|u}$ for $u>0$
has $K_{j,u}$ codewords, each consisting of $n_{j,u}$ symbols $\tilde{x}_{j,u,l}\in\{0,1\}$,
which are randomly and independently generated as $\verb"Bern"(p_{j|u})$
variables, with $l=1,2,...,n_{j,u}$ and $n_{j,u}=n\delta_{j,u}$,
for some $0\leq\delta_{j,u}<1$. We denote the codewords as $\tilde{x}_{j,u}^{n_{j,u}}(m_{j,u})$
with $m_{j,u}\in[1,K_{j,u}]$. Note that the parameter $\delta_{j,u}$
does not depend on $n$, and hence, if $n\rightarrow\infty$, then
we have $n_{j,u}\rightarrow\infty$ for all $j,u$. We set $2^{nR_{j}}=\prod_{u=1}^{\verb"U"}K_{j,u}$,
while the relations among the remaining parameters ($K_{j,u}$,$\delta_{j,u}$,$p_{j|u}$)
will be specified below.


\subsubsection{Encoding}

Each node performs rate splitting. Namely, given a message $M_{j}\in[1,2^{nR_{j}}]$,
Node $j$ finds a $\verb"U"-$tuple $[m_{j,1},...,m_{j,\verb"U"}]$
with $m_{j,u}\in[1,K_{j,u}]$ that uniquely represents $M_{j}$. This
is always possible since we have $2^{nR_{j}}=\prod_{u=1}^{\verb"U"}K_{j,u}$.
Then, the selected codewords $\tilde{x}_{j,u}^{n_{j,u}}(m_{j,u})$
for $u\in[1,\verb"U"]$ are transmitted via multiplexing based on
the current available energy. Specifically, each Node $j$ initializes
$\verb"U"$ pointers $l_{j,1}=l_{j,2}=\cdots=l_{j,\verb"U"}=1$ that
keep track of the number of symbols already sent from codewords $\tilde{x}_{j,1}^{n_{j,1}}(m_{j,1})$,
$\tilde{x}_{j,2}^{n_{j,2}}(m_{j,2})$, ..., $\tilde{x}_{j,\verb"U"}^{n_{j,\verb"U"}}(m_{j,\verb"U"})$,
respectively. At channel use $i$, if the state is $U_{1,i}=u$, then
the nodes operate as follows. 
\begin{itemize}
\item Node 1: If $u=0$, then $x_{1,i}=0$. Else, if $l_{1,u}\leq n_{1,u}$,
Node $1$ transmits $x_{1,i}=\tilde{x}_{1,u,l_{1,u}}(m_{1,u})$ and
increments the pointer $l_{1,u}$ by 1. Finally, if $l_{1,u}=n_{1,u}+1$
the pointer $v_{1,u}$ is not incremented, and the transmitter uses
\emph{random padding}, i.e., it sends $x_{1,i}=1$ with probability
$p_{1,u}$ and $x_{1,i}=0$ otherwise. 
\item Node 2: If $u=\verb"U"$ (i.e., no energy is available at Node 2),
then $x_{2,i}=0$. Else, if $l_{2,\verb"U"-u}\leq n_{2,\verb"U"-u}$,
Node $2$ transmits $x_{2,i}=\tilde{x}_{2,\verb"U"-u,2,l_{2,\verb"U"-u}}(m_{2,\verb"U"-u})$
and increments the pointer $l_{2,\verb"U"-u}$ by 1. Finally, if $l_{2,\verb"U"-u}=n_{2,\verb"U"-u}+1$,
the pointer $l_{2,\verb"U"-u}$ is not incremented, and Node 2 sends
$x_{2,i}=1$ with probability $p_{2,\verb"U"-u}$ and $x_{2,i}=0$
otherwise. 
\end{itemize}
The random padding method used above is done for technical reasons
that will be clarified below.

\subsubsection{Decoding}

We first describe the decoding strategy for Node 2. By construction,
the nodes are aware of the state sequence $U_{1}^{n}$, and thus can
determine the ordered set 
\begin{equation}
{\cal N}_{u}=\{i|U_{1,i}=u\},
\end{equation}
 of channel use indices in which the state is $u$ with $u\in[0,$$\verb"U"${]}.
For all $u\in[1,$$\verb"U"${]}, if $|{\cal N}_{u}|\geq n_{1,u}$,
then Node 2 takes the first $n_{1,u}$ indices $i_{u,1}<i_{u,2}<\cdots<i_{u,n_{1,u}}$
from the set ${\cal N}_{u}$ and obtains the list of messages $m_{1,u}\in[1,K_{1,u}]$
that satisfy $\tilde{x}_{1,u,k}(m_{1,u})=x_{1,i_{u,k}}$ for all $k\in[1,n_{1,u}]$.
Note that the list cannot be empty due to the fact that the channel
is noiseless. However, it contains more than one message, or if $|\mathcal{N}_{u}|<n_{1,u}$,
then Node 2 puts out the estimate $\hat{m}_{1,u}=1$. Instead, if
the list contains only one message $m_{1,u}$, then Node 2 sets $\hat{m}_{1,u}=m_{1,u}$.
Finally, the message estimate is obtained as $\hat{m}_{1}=[\hat{m}_{1,1},...,\hat{m}_{1,\verb"U"}]$.

Node 1 operates in the same way, with the only caveat that the $u$th
codebook $\mathcal{C}_{2|u}$ of Node 2 is observed at channel uses
in the set ${\cal N}_{U-u}$ for $u\in[1,\verb"U"]$.

\subsubsection{Analysis}

We evaluate the probability of error on average over the messages
and the generation of the codebooks, following the random coding principle.
From the definition of the decoders given above, the event that any
of the decoders is in error is included in the set ${\cal E}=\bigcup_{j=1,2}\bigcup_{u=1}^{\verb"U"}({\cal E}_{j,u}^{(1)}\cup{\cal E}_{j,u}^{(2)})$,
where: (\emph{i}) ${\cal E}_{j,u}^{(1)}$ is the event that $|\mathcal{N}_{u}|<n_{1,u}$
for $j=1$ and that $|\mathcal{N}_{\verb"U"-u}|<n_{2,u}$ for $j=2$,
that is, that the number of channel uses in which the system resides
in the state in which the codeword $\tilde{x}_{j,u}^{n_{j,u}}(m_{j,u})$
from the codebook $\mathcal{C}_{j,u}$ is sent is not sufficient to
transmit the codeword in full; (\emph{ii}) ${\cal E}_{j,u}^{(2)}$
is the event that two different messages $m_{j,u}^{\prime},m_{j,u}^{\prime\prime}\in[1,K_{j,u}]$
are represented by the same codewords, i.e., $\tilde{x}_{j,u}^{n_{1,u}}(m_{j,u}^{\prime})=\tilde{x}_{1,u}^{n_{,1u}}(m_{1,u}^{\prime\prime})$.

The probability of error can thus be upper bounded as 
\begin{equation}
\Pr[\mathcal{E}]\leq\sum_{j=1}^{2}\sum_{u=1}^{\verb"U"}\left(\Pr[{\cal E}_{j,u}^{(1)}]+\Pr[{\cal E}_{j,u}^{(2)}]\right).\label{eq:Pje}
\end{equation}
 In the following, we evaluate upper bounds on this terms.

It immediately follows from the packing lemma of \cite{El Gamal Kim}
that $\Pr[{\cal E}_{j,u}^{(2)}]\rightarrow0$ as $n_{j,u}\rightarrow\infty$
as long as 
\begin{equation}
\frac{\log_{2}K_{j,u}}{n_{j,u}}<H(p_{j|u})-\delta(\epsilon)
\end{equation}
 with $\delta(\epsilon)\rightarrow0$ as $\epsilon\rightarrow0$.
For analysis of the probabilities $\Pr[{\cal E}_{j,u}^{(1)}]$, we
observe that, under the probability measure induced by the described
random codes, the evolution of the state $U_{1,i}$ across the channel
uses $i\in[1,n]$ is a Markov chain with $\verb"U"+1$ states. Specifically,
the chain is a birth-death process, since, if the state is $U_{1,i}=u$
in channel use $i$, the next state $U_{1,i+1}$ can only be either
$u-1$ or $u+1$. More precisely, let $q_{u|w}=\Pr(U_{1,i+1}=u|U_{1,i}=w$)
be the transition probability. Note that, due to the use of random
padding, the transition probability $q_{u|w}$ remains constant during
all $n$ channel uses, so that the Markov chain is time-invariant.

We now elaborate on the Markov chain $U_{1,i}$. To this end, we first
define as $\phi_{x_{1},x_{2}|u}$, where $x_{1},x_{2}\in\{0,1\}$
be the joint probability that Node 1 transmits $X_{1,i}=x_{1}$ and
Node 2 transmits $X_{2,i}=x_{2}$ during the $i$th channel use in
which the state is $U_{1,i}=u$. We can now write the non-zero values
of the transition probability $q_{u|w}$ as follows: 
\begin{align}
q_{u,u-1}=\phi_{1,0|u}\quad q_{u,u+1}=\phi_{0,1|u}\nonumber \\
q_{u,u}=1-q_{u,u-1}-q_{u,u+1}
\end{align}
 With a slight abuse of the notation and noting that $\phi_{1,0|0}=\phi_{1,1|0}=0$
and $\phi_{0,1|\verb"U"}=\phi_{1,1|\verb"U"}=0$ the expressions above
also represent the transitions for the two extremal states $u=0$
and $u=\verb"U"$, as they imply $q_{0|-1}=0$ and $q_{\verb"U"|\verb"U"+1}=0$.

If $p_{1,0}=p_{2,0}=0$ and $0<p_{1,u},p_{2,u}<1$ for all $u>0$,
then it can be seen that the Markov chain is aperiodic and irreducible,
and thus there exist a unique set of stationary probabilities $\pi_{0},\pi_{1},\cdots,\pi_{\verb"U"}$,
which are given by solving the linear system, defined by taking \verb"U"
equations of type (\ref{eq:MarkovChainTransitions}) for $u=0\ldots\verb"U"-1$
and adding the condition $\sum_{u=0}^{\verb"U"}\pi_{u}=1$.

We are now interested in the statistical properties of the set $|\mathcal{N}_{u}|$
of channel uses in which the state satisfies $U_{1}=u$. Using the
ergodic theorem and the strong law of large numbers \cite[Theorem 1.10.2]{Norris},
it can be shown that $\lim_{n\rightarrow\infty}\frac{V_{u}(n)}{n}=\pi_{u}$
with probability 1. Therefore, if we choose: 
\begin{equation}
l_{1,u}=l_{2,\verb"U"-u}=n(\pi_{u}-\epsilon)
\end{equation}
 then $\Pr[{\cal E}_{1,u}^{(2)}]=\Pr[{\cal E}_{2,\verb"U"-u}^{(2)}]$
can be made arbitrarily close to $0$ as $n\rightarrow\infty$. This
concludes the proof.

\section{Proof of Proposition 2}

Consider any $(n,R_{1},R_{2},\verb"U")$ code with zero probability
of error, as per our definition of achievability in Sec. \ref{sec:SysModel}.
We have the following inequalities: 
\begin{align}
nR_{1} & =H(M_{1})=H(M_{1}|M_{2},U_{1,1}=u_{1,1})\nonumber \\
 & \overset{\textrm{\ensuremath{(a)}}}{=}H(M_{1},X_{1}^{n},U_{1}^{n}|M_{2},U_{1,1}=u_{1,1})\nonumber \\
 & \overset{(b)}{=}H(X_{1}^{n},U_{1}^{n}|M_{2},U_{1,1}=u_{1,1})\nonumber \\
 & \overset{}{=}\sum_{i=1}^{n}H(X_{1,i},U_{1,i}|X_{1}^{i-1},U_{1}^{i-1},M_{2},U_{1,1}=u_{1,1})\nonumber \\
 & =\sum_{i=1}^{n}H(U_{1,i}|X_{1}^{i-1},U_{1}^{i-1},M_{2},U_{1,1}=u_{1,1})\nonumber \\
 & +H(X_{1,i}|X_{1}^{i-1},U_{1}^{i},M_{2},U_{1,1}=u_{1,1})\nonumber \\
 & \overset{(c)}{=}\sum_{i=1}^{n}H(X_{1,i}|X_{1}^{i-1},U_{1}^{i},M_{2},U_{1,1}=u_{1,1})\nonumber \\
 & \overset{(d)}{\leq}\sum_{i=1}^{n}H(X_{1,i}|U_{1,i},X_{2,i})\nonumber \\
 & \overset{(e)}{=}H(X_{1}|U_{1},X_{2},Q)\nonumber \\
 & \leq H(X_{1}|U_{1},X_{2}),
\end{align}
 where (\emph{a}) follows since $X_{1}^{n},U_{1}^{n}$ are functions
of $M_{1},M_{2}$ and $u_{1,1}$; (\emph{b}) follows since $H(M_{1}|X_{1}^{n},U_{1}^{n},$
$M_{2},U_{1,1}=u_{1,1})=0$ holds due to the constraint of zero probability
of error; (\emph{c}) follows since $U_{1,i}$ is a function of $X_{1}^{i-1},M_{2}$
and $u_{1,1}$; (\emph{d}) follows by conditioning reduces entropy;
(e) follows by defining a variable $Q$ uniformly distributed in the
set $[1,n]$ and independent of all other variables, along with $X_{1}=X_{1Q}$,
$X_{2}=X_{2Q}$ and $U_{1}=U_{1Q}$.

Similar for $nR_{2}$ we obtain the bound $nR_{1}\leq H(X_{1}|U_{1},X_{2})$.
Moreover, for the sum-rate, similar steps lead to 
\begin{align}
n(R_{1}+R_{2}) & =H(M_{1},M_{2})=H(M_{1},M_{2}|U_{1,1}=u_{1,1})\nonumber \\
 & =H(M_{1}M_{2},X_{1}^{n},X_{2}^{n},U_{1}^{n}|U_{1,1}=u_{1,1})\nonumber \\
 & =H(X_{1}^{n},X_{2}^{n},U_{1}^{n}|U_{1,1}=u_{1,1})\nonumber \\
 & =\sum_{i=1}^{n}H(U_{1,i}|X_{1}^{i-1},X_{2}^{i-1},U_{1}^{i-1},M_{2},U_{1,1}=u_{1,1})\nonumber \\
 & +H(X_{1,i},X_{2,i}|X_{1}^{i-1},X_{2}^{i-1},U_{1}^{i},M_{2},U_{1,1}=u_{1,1})\nonumber \\
 & \geq\sum_{i=1}^{n}H(X_{1,i},X_{2,i}|U_{1,i})\nonumber \\
 & =H(X_{1},X_{2}|U_{1}).
\end{align}

Let us now define $\pi_{u}=\Pr[U_{1}=u]$ and $\phi_{x_{1},x_{2}|u}=\Pr[X_{1}=x_{1},X_{2}=x_{2}|U_{1}=u]$
for $i,j\in\{0,1\}$ and for all $u_{1}\in\{0,1,...,\verb"U"\}$.
Probability conservation implies that the relationship (\ref{eq:MarkovChainTransitions})
be satisfied. This concludes the proof.

\section{Transition Probabilities for the Model in Sec. \ref{sec:Generalizing-the-Main}\label{app:C}}

\begin{figure}[h!]
\centering \includegraphics[clip,scale=0.3]{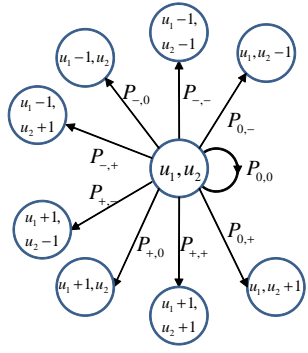} \caption{The outgoing transition probabilities from a state $(u_{1},u_{2})$.}

\label{fig:fig4} 
\end{figure}

Here we discuss the transition probability matrix $\boldsymbol{P}$
used in Sec. \ref{sec:Generalizing-the-Main}. To this end, define
as $Q_{ij}^{(ab)}$ ($i,j=1,2$) the probability that $H_{j}=a\in\{0,1\}$
energy units are added to the battery at Node $j$ conditioned on
Node $i$ sending symbol $X_{j}=b\in\{0,1\}$, for $i\neq j$, namely
\begin{subequations} 
\begin{eqnarray}
Q_{ij}^{(00)} & = & P_{ij}^{(00)}P_{jj}^{(00)}+P_{ij}^{(01)}P_{jj}^{(10)},\\
Q_{ij}^{(01)} & = & P_{ij}^{(00)}P_{jj}^{(01)}+P_{ij}^{(01)}P_{jj}^{(11)},\\
Q_{ij}^{(10)} & = & P_{ij}^{(10)}P_{jj}^{(00)}+P_{ij}^{(11)}P_{jj}^{(10)},\\
\textrm{and }Q_{ij}^{(11)} & = & P_{ij}^{(10)}P_{jj}^{(01)}+P_{ij}^{(11)}P_{jj}^{(11)}.
\end{eqnarray}
 \end{subequations} Note that these transition probabilities correspond
to the cascade of the channels in Fig \ref{fig:fig2} and Fig \ref{fig:fig3}.
Based on these probabilities, we can now evaluate all the possible
transition probabilities from state $(u_{1},u_{2})$ to any other
state $(u_{1}^{\prime},u_{2}^{\prime})$. We start with $u_{1}\in[1,B_{1}-1]$
and $u_{2}\in[1,B_{2}-1]$ for $B_{1},B_{2}>1$ whose outgoing transition
probabilities are illustrated in Fig. \ref{fig:fig4}. The ``boundary''
states with $u_{j}=0$ or $u_{j}=B_{j}$ for some $j=1,2$ are discussed
later. 

By the stated assumptions, the state $(u_{1},u_{2})$ can only transit
to state $(u_{1}+i_{1},u_{2}+i_{2})$ with $i_{1},i_{2}\in\{-1,0,1\}$,
so that the energy in the battery is increased or decreased by at
most one energy unit. Therefore, for the ``non-boundary'' states
$(u_{1},u_{2})$ with $u_{1}\in[1,B_{1}-1]$ and $u_{2}\in[1,B_{2}-1]$,
the probabilities in Fig. \ref{fig:fig4} can be easily obtained as
\begin{eqnarray}
P_{0,0}(u_{1},u_{2}) & = & p_{1|(u_{1},u_{2})}p_{2|(u_{1},u_{2})}(Q_{21}^{(11)}Q_{12}^{(11)})+p_{1|(u_{1},u_{2})}\bar{p}_{2|(u_{1},u_{2})}(Q_{21}^{(01)}Q_{12}^{(10)})\nonumber \\
 &  & +\bar{p}_{1|(u_{1},u_{2})}p_{2|(u_{1},u_{2})}(Q_{21}^{(10)}Q_{12}^{(01)})+\bar{p}_{1|(u_{1},u_{2})}\bar{p}_{2|(u_{1},u_{2})}(Q_{21}^{(00)}Q_{12}^{(00)}),\label{eq:p00}
\end{eqnarray}

\begin{eqnarray}
P_{+,0}(u_{1},u_{2}) & = & \bar{p}_{1|(u_{1},u_{2})}p_{2|(u_{1},u_{2})}(Q_{21}^{(11)}Q_{12}^{(01)})+\bar{p}_{1|(u_{1},u_{2})}\bar{p}_{2|(u_{1},u_{2})}(Q_{21}^{(01)}Q_{12}^{(00)}),\label{eq:p+0}
\end{eqnarray}

\begin{eqnarray}
P_{0,+}(u_{1},u_{2}) & = & p_{1|(u_{1},u_{2})}\bar{p}_{2|(u_{1},u_{2})}(Q_{21}^{(01)}Q_{12}^{(11)})+\bar{p}_{1|(u_{1},u_{2})}\bar{p}_{2|(u_{1},u_{2})}(Q_{21}^{(00)}Q_{12}^{(01)}),\label{eq:p0+}
\end{eqnarray}

\begin{eqnarray}
P_{+,+}(u_{1},u_{2}) & = & \bar{p}_{1|(u_{1},u_{2})}\bar{p}_{2|(u_{1},u_{2})}(Q_{21}^{(01)}Q_{12}^{(01)}),\label{eq:p++}
\end{eqnarray}

\begin{eqnarray}
P_{+,-}(u_{1},u_{2}) & = & \bar{p}_{1|(u_{1},u_{2})}p_{2|(u_{1},u_{2})}(Q_{21}^{(11)}Q_{12}^{(00)}),\label{eq:p+-}
\end{eqnarray}

\begin{eqnarray}
P_{-,+}(u_{1},u_{2}) & = & p_{1|(u_{1},u_{2})}\bar{p}_{2|(u_{1},u_{2})}(Q_{21}^{(00)}Q_{12}^{(11)}),\label{eq:p-+}
\end{eqnarray}

\begin{eqnarray}
P_{0,-}(u_{1},u_{2}) & = & p_{1|(u_{1},u_{2})}p_{2|(u_{1},u_{2})}(Q_{21}^{(11)}Q_{12}^{(10)})+\bar{p}_{1|(u_{1},u_{2})}p_{2|(u_{1},u_{2})}(Q_{21}^{(10)}Q_{12}^{(00)}),\label{eq:p0-}
\end{eqnarray}

\begin{eqnarray}
P_{-,0}(u_{1},u_{2}) & = & p_{1|(u_{1},u_{2})}p_{2|(u_{1},u_{2})}(Q_{21}^{(10)}Q_{12}^{(11)})+p_{1|(u_{1},u_{2})}\bar{p}_{2|(u_{1},u_{2})}(Q_{21}^{(00)}Q_{12}^{(10)}),\label{eq:p-0}
\end{eqnarray}
 and 
\begin{eqnarray}
P_{-,-}(u_{1},u_{2}) & = & p_{1|(u_{1},u_{2})}p_{2|(u_{1},u_{2})}(Q_{21}^{(10)}Q_{12}^{(10)}),\label{eq:p--}
\end{eqnarray}
 where we recall that $p_{i|(u_{1},u_{2})}$ is the probability of
sending a ``$1$'' symbol by Node $i$ given the state $(u_{1},u_{2})$
and $\bar{p}_{i|(u_{1},u_{2})}=1-p_{i|(u_{1},u_{2})}$.

For the \textquotedbl{}boundary\textquotedbl{} states $(u_{1},u_{2})$
with $u_{1}$ and/or $u_{2}$ equal to 0 the outgoing transitions
in Fig. \ref{fig:fig4} and probabilities in (\ref{eq:p00})-(\ref{eq:p--})
still hold since the transitions to states with energy less than zero
are dis enabled by the conditions $p_{i|(u_{1},u_{2})}=0$ for $u_{j}=0$,
$j=1,2$. Instead, if $u_{1}=B_{1}$ and $u_{2}\in[0,B_{2}-1]$ then
the probabilities in (\ref{eq:p0-})-(\ref{eq:p--}) remain the same,
but we have $P_{+,-}=P_{+,0}=P_{+,+}=0$ and $P_{0,0}$ equals the
sum of the right-hand sides of (\ref{eq:p00}) and (\ref{eq:p+0}),
while $P_{0,+}$ equals the sum of the right-hand sides of (\ref{eq:p0+})
and (\ref{eq:p++}), and $P_{0,-}$ equals the sum of the right-hand
sides of (\ref{eq:p0-}) and (\ref{eq:p+-}) . The transition probabilities
from the states with $u_{1}\in[0,B_{1}-1]$ and $u_{2}=B_{2}$ follow
in a symmetric fashion. Finally if $u_{1}=B_{1}$ and $u_{2}=B_{2}$,
then we have $P_{-,+}=P_{+,-}=P_{+,0}=P_{0,+}=P_{+,+}=0$ and $P_{0,0}$
is the sum of (\ref{eq:p00}), (\ref{eq:p++}), (\ref{eq:p+0}), and
(\ref{eq:p0+}), while $P_{0,-}$ is the sum of (\ref{eq:p0-}) and
(\ref{eq:p+-}) $\textrm{and }P_{-,0}$ is the sum of (\ref{eq:p-0})
and (\ref{eq:p-+}).

By using the transition probabilities defined above, one can easily
construct the transition matrix $\boldsymbol{P}$ of the corresponding
Markov chain. For instance, for the case $B=1$, we can write the
transition matrix as 
\begin{eqnarray}
\boldsymbol{P}\negmedspace\negmedspace & \negmedspace\negmedspace=\negmedspace\negmedspace & \negmedspace\negmedspace\left[\negmedspace\negmedspace\negmedspace\begin{array}{cccc}
P_{0,0}(0,0)\negmedspace\negmedspace\negmedspace & P_{0,-}(0,1)\negmedspace\negmedspace\negmedspace & P_{-,0}(1,0)\negmedspace\negmedspace\negmedspace & P_{-,-}(1,1)\\
P_{0,+}(0,0)\negmedspace\negmedspace\negmedspace & P_{0,0}(0,1)\negmedspace\negmedspace\negmedspace & P_{-,+}(1,0)\negmedspace\negmedspace\negmedspace & P_{-,0}(1,1)\\
P_{+,0}(0,0)\negmedspace\negmedspace\negmedspace & P_{+,-}(0,1)\negmedspace\negmedspace\negmedspace & P_{0,0}(1,0)\negmedspace\negmedspace\negmedspace & P_{0,-}(1,1)\\
P_{+,+}(0,0)\negmedspace\negmedspace\negmedspace & P_{+,0}(0,1)\negmedspace\negmedspace\negmedspace & P_{0,+}(1,0)\negmedspace\negmedspace\negmedspace & P_{0,0}(1,1)
\end{array}\negmedspace\negmedspace\negmedspace\right],
\end{eqnarray}
 where the column index represents the the initial state and the row
index the final state.

\end{document}